\documentclass[sigconf, nonacm, pdfa]{acmart}

\newcommand\vldbdoi{10.14778/3681954.3681961}
\newcommand\vldbpages{2764 - 2777}
\newcommand\vldbvolume{17}
\newcommand\vldbissue{11}
\newcommand\vldbyear{2024}
\newcommand\vldbauthors{\authors}
\newcommand\vldbtitle{\shorttitle} 
\newcommand\vldbavailabilityurl{https://github.com/Rustam-Warwick/d3-gnn}
\newcommand\vldbpagestyle{empty} 
\newcommand{\Ours}{D3-GNN}

\usepackage{subcaption}
\usepackage{caption}
\usepackage[a-2b]{pdfx}
\usepackage{hyperref}
\usepackage{balance}
\usepackage[htt]{hyphenat}
\usepackage{comment}
\usepackage{enumitem}
\usepackage{todonotes}
\usepackage{multirow}
\usepackage{algorithm}
\usepackage[noend]{algpseudocode}
\usepackage{amsmath}
\usepackage[normalem]{ulem}

\usepackage{xcolor}
\usepackage{framed}
\definecolor{shadecolor}{RGB}{235,235,235}

\algnewcommand{\IfThen}[2]{\State \algorithmicif\ #1\ \algorithmicthen\ #2}
\algnewcommand{\ForAllDo}[2]{\State \algorithmicforall\ #1\ \algorithmicdo\ #2}
\algnewcommand{\WhileDo}[2]{\State \algorithmicwhile\ #1\ \algorithmicdo\ #2}

\begin{document}

\title{D3-GNN: Dynamic Distributed Dataflow for Streaming Graph
Neural Networks}

\author{Rustam Guliyev}
\affiliation{%
  \institution{University of Warwick}
  \city{Coventry}
  \country{UK}
}
\email{rustam.guliyev@warwick.ac.uk}

\author{Aparajita Haldar}
\authornote{Currently with Fujitsu Research of Europe. This publication describes work performed at the University of Warwick and is not associated with Fujitsu.}
\affiliation{%
  \institution{University of Warwick}
  \city{Coventry}
  \country{UK}
}
\email{aparajita.haldar@warwick.ac.uk}

\author{Hakan Ferhatosmanoglu}
\authornote{Also with Amazon Web Services. This publication describes work performed at the University of Warwick and is not associated with Amazon.}
\affiliation{%
  \institution{University of Warwick}
  \city{Coventry}
  \country{UK}
}
\email{hakan.f@warwick.ac.uk}

\begin{abstract}
Graph Neural Network (GNN) models on streaming graphs entail
algorithmic challenges to continuously capture its dynamic state, as well as systems challenges to optimize latency, memory, and throughput during both inference and training. We present {\Ours}, the first distributed, hybrid-parallel, streaming GNN system designed to handle real-time graph updates under online query setting. Our system addresses data management, algorithmic, and systems challenges, enabling continuous capturing of the dynamic state of the graph and updating node representations with fault-tolerance and optimal latency, load-balance, and throughput. {\Ours} utilizes streaming GNN aggregators and an unrolled, distributed computation graph architecture to handle cascading graph updates. To counteract data skew and neighborhood explosion issues, we introduce inter-layer and intra-layer windowed forward pass solutions. Experiments on large-scale graph streams demonstrate that {\Ours} achieves high efficiency and scalability. Compared to DGL, {\Ours} achieves a significant throughput improvement of about 76x for streaming workloads. The windowed enhancement further reduces running times by around 10x and message volumes by up to 15x at higher parallelism.
\end{abstract}

\maketitle

\pagestyle{\vldbpagestyle}
\begingroup\small\noindent\raggedright\textbf{PVLDB Reference Format:}\\
\vldbauthors. \vldbtitle. PVLDB, \vldbvolume(\vldbissue): \vldbpages, \vldbyear.\\
\href{https://doi.org/\vldbdoi}{doi:\vldbdoi}
\endgroup
\begingroup
\renewcommand\thefootnote{}\footnote{\noindent
This work is licensed under the Creative Commons BY-NC-ND 4.0 International License. Visit \url{https://creativecommons.org/licenses/by-nc-nd/4.0/} to view a copy of this license. For any use beyond those covered by this license, obtain permission by emailing \href{mailto:info@vldb.org}{info@vldb.org}. Copyright is held by the owner/author(s). Publication rights licensed to the VLDB Endowment. \\
\raggedright Proceedings of the VLDB Endowment, Vol. \vldbvolume, No. \vldbissue\ %
ISSN 2150-8097. \\
\href{https://doi.org/\vldbdoi}{doi:\vldbdoi} \\
}\addtocounter{footnote}{-1}\endgroup

\ifdefempty{\vldbavailabilityurl}{}{
\vspace{.3cm}
\begingroup\small\noindent\raggedright\textbf{PVLDB Artifact Availability:}\\
The source code, data, and/or other artifacts have been made available at \url{\vldbavailabilityurl}.
\endgroup
}
\section{Introduction}

The ubiquity of large-scale, {semi-structured} data, such as knowledge graphs, social networks, financial transactions, and e-commerce networks, has fostered graph learning~\cite{sahu2017ubiquity}. To this end, Graph Neural Networks (GNNs) have proven to achieve greater performance on tasks like node classification~\cite{hamilton2017inductive}, link prediction~\cite{zhan2018linkprediction}, and graph classification~\cite{feng2021hierarchical}, compared to traditional approaches~\cite{xu2018how}. GNNs combine neural networks (NN) with graph topology, allowing them to generate semantically richer node embeddings used for downstream prediction tasks. Applications of these models and the various tasks have since been extensively studied in recommendation systems~\cite{yang2021consisrec, gao2023surveygnnrecommender}, computer vision~\cite{han2022vision, bera2022srgnn}, social networks~\cite{ijcai2018p142}, fraud detection~\cite{lu20222bright}, and more.

Several frameworks have been developed to facilitate distributed GNN model development and deployment in static settings~\cite{vatter2023@evolutionofdistributedgnn}. However, most of the graphs observed in the real world are dynamic or streaming~\cite{chiang@2019clustergcn}. For example, in social networks, new users join and existing users may update their profiles or interests, leading to node updates and additions. The rate of ingestion in a streaming graph can be high, such as 30K edges/sec in Alibaba graph~\cite{qiu2018alibaba}. In addition to streaming updates, it is also important to consider low-latency query settings in the design of GNN systems. The state-of-the-art methods typically follow the \textbf{ad-hoc} querying setting, where an external actor is expected to initiate the execution of a GNN query. 

However, a rather unexplored setting, which is of high importance for latency-critical applications, is the \textbf{online} setting where queries are implicit to graph topology. Consider a dynamic social network where the system needs to autonomously identify and monitor potential spam accounts or harmful content creators without external prompt. This requires the GNN system to continuously analyze the graph for such behaviors, a task that cannot be efficiently handled by ad-hoc systems due to their reliance on periodic entire-graph inferences, thus making them ill-suited for near-real-time requirements. We bridge these gaps by introducing a distributed, end-to-end solution that is highly flexible and is designed to handle streaming graphs in online query settings.

Under streaming graph updates, previously generated representations become stale and require cascading GNN inference operations over a set of influenced nodes to maintain an up-to-date state. During inference (forward pass), the cascading nature of operations causes \textit{neighborhood explosion}~\cite{chiang@2019clustergcn}, which makes it challenging to maintain consistent throughput while keeping representations abreast of the updates. Identifying the set of influenced nodes (i.e., nodes affected by graph updates) is also resource-intensive, requiring a full L-hop traversal of the out-neighborhood. The irregular access patterns and data-skew on graphs also incur additional overheads, due to central (hub) nodes being involved in the majority of computations. This worsens the problems caused by 'neighborhood explosion' and introduces significant load-imbalance. Also, updates can incur \textit{concept drifts} over time, which deems periodic model re-training (backward pass) necessary to sustain model accuracy. All these challenges have to be tackled while keeping in mind the scalability and fault-tolerance of the system, which becomes more challenging in high-volume, streaming workloads. Hence, as the system requirements get closer to near-real-time, it becomes increasingly difficult to manage its latency, memory, and throughput constraints using traditional approaches.

Nowadays, we have libraries for GNN processing like DGL~\cite{wang2019deep}, Pytorch Geometric~\cite{fast_graph_representation} and Pytorch Geometric Temporal~\footnote{https://pytorch-geometric-temporal.readthedocs.io/}. These provide APIs for storing, manipulating and performing spatial operations on graphs. However, under the streaming workloads of large graphs, these fall short due to their inability to utilize a distributed environment.
By contrast, the few libraries for distributed GNNs (e.g., DistDGL~\cite{zheng2020distdgl}, AliGraph~\cite{zhu2019aligraph}, DynaGraph~\cite{dynagraph}) are not built for streaming graphs or online queries. Some recent work considers temporal GNN models~\cite{zhou2022tgl, huan2023tgcn}, however, they all treat the graph as a static one with temporal edges and are mostly designed only for training workloads. These employ static data structures and static graph partitioning algorithms as a pre-processing step to distribute the graphs. For processing, they follow a synchronous, mini-batch execution model which requires transferring ego-graphs and their raw features to distributed machines to then perform local GNN computations. This violates the data locality principle and forces graph data to be migrated redundantly with every batch iteration, which can take up to 85\% of the system's compute time~\cite{lin2022characterizing}. Furthermore, no prior system considers the impact of external data-skews on GNN workloads. Retrofitting these features into such systems introduces further maintainability problems, and delayed predictions, rendering them unsuitable for latency-critical applications. Hence, there is a need for a holistic GNN system that incrementally operates on large graph updates while also efficiently tackling the \textbf{online} query settings.

To address the above challenges, we present {\Ours} - the first distributed and hybrid-parallel dataflow-based system for streaming GNNs. {\Ours} enables asynchronous, incremental GNN inference within a scalable dataflow pipeline. As  node representations are fundamentally the building block for virtually all graph-learning tasks, the {\Ours} system is designed to maintain up-to-date node representations under streaming graph updates in an online fashion. This core capability allows {\Ours} to be versatile across different models and applications, in other words fundamentally task-agnostic. To build {\Ours}, we leverage Apache Flink\footnote{\url{https://flink.apache.org}}, which helps us to develop an extendable and fault-tolerant system and tackle intrinsic challenges in event streams, such as exactly-once processing and handling late events.

At its heart, {\Ours} employs a novel design of unrolled computation graph, with each dataflow operator representing a GNN layer, facilitating both data and model parallelism via streaming graph partitioning and the separation of GNN layers across operators, respectively. To efficiently handle streaming graph updates, we pivot on node memory and streaming aggregators, which perform consistent, incremental updates in node states. Furthermore, we propose and analyze several windowing algorithms for tackling the neighborhood explosion and data-skew issues persistent in GNNs with only a minor latency trade-off. Lastly, while continual learning algorithms for GNNs is an orthogonal research problem, providing an effective system support for it in an inference-first pipeline solves commonly occurring problems of model staleness, data migration, and the need for periodic resource provisioning. Hence, although {\Ours} is primarily designed as an inference-first system, it also supports efficient distributed training.

We further enhance our system by solutions for efficient graph \& model storage, stale-free training, feature replication, tensor management, nested iterations, and termination detection. The decoupled nature  of {\Ours} allows independent parallelization of GNN layers and flexibility in modifying the dataflow egress depending on the specific use-case. For example, streaming node representations can be used to generate real-time predictions or act as a materialized embedding table that can be further queried. {\Ours}'s model-agnostic approach, based on MPGNN, enables support for a wide range of common GNN models.

\section{Related Work}
\label{section::realted_work}
To the best of our knowledge, there is no other streaming dataflow solution for GNN computations tackling the online query execution model, either as a research or as an industrial offering. Existing systems are not well-suited for low-latency streaming graph learning and inference operations.

\subsection{Streaming Graph Processing Systems}
Over the past decade, a variety of graph management systems and algorithms have been designed for streaming graph analytics~\cite{mondal2012managing,choudhury2013streamworks,pacaci2020regular,tian2023world}. These systems are optimized for running analytics algorithms, such as subgraph matching and subgraph counting, on such dynamic data~\cite{mcgregor2014graph}. However, none of these systems are designed to perform on graph learning tasks such as GNNs~\cite{erb2018graphtides}.

Recently, with the advent of streaming systems, dataflow-based graph systems have been developed. The Gelly-streaming\footnote{https://github.com/vasia/gelly-streaming} library, for instance, built on top of Apache Flink, tackles streaming graph algorithms such as finding connected components and bipartite matching. GraphX~\cite{gonzalez2014graphx} aims to unify optimizations from specialized graph processing environments (e.g., Pregel) with general-purpose processing environments (e.g., MapReduce, Apache Spark). Other dataflow-based methods have been devised for incremental graph algorithms, such as Naiad~\cite{murray2013naiad}, GraphTau~\cite{iyer2016time}, Tornado~\cite{shi2016tornado}, and KickStarter~\cite{vora2017kickstarter}. None of these tackle graph learning and GNN computations in the streaming setting.

By contrast, {\Ours} provides a dataflow-based streaming graph system  targeting GNN computations and applications. It enables incremental GNN computations while preserving primary aspects of streaming systems~\cite{carbone2020beyond}. Note that studies on enhancing dynamic GNN training (e.g., edge events, continual learning, memory modules~\cite{nguyen2018continuous,ma2020streaming,wang2020streaming,rossi2020temporal}) are orthogonal to our work, and most can be implemented within our modular {\Ours} ecosystem.

\subsection{Distributed GNN Systems}
Extending from graph analytics systems, the recent advent of GNNs has brought attention to push-based distributed systems. These aim to provide data/model-parallel ML capabilities to message-passing models~\cite{nagrecha2021model,demirci2021partitioning}. NeuGraph~\cite{ma2019neugraph} proposed an abstraction for expressing GNN pipelines with graph-specific optimizations. Here, backward gather functions are used to distribute the accumulated gradients on the backward pass, thereby supporting distributed parallel GNN training. {\Ours} proposes a streaming aggregator perspective with the help of a dynamic dataflow middleware. 

To overcome communication overheads in large-scale training, sampling methods such as neighborhood sampling (e.g., GraphSAGE~\cite{hamilton2017inductive}) have been proposed. Motivated by this, recent developments of GNN systems like AliGraph~\cite{zhu2019aligraph}, DistDGL~\cite{zheng2020distdgl}, DistDGLv2~\cite{zheng2021distdglv2}, AGL~\cite{zhang2020agl}, and FlexGraph~\cite{wang2021flexgraph} employ distributed, mini-batch sampling operators to perform pull-based GNN computations. These collect (pull) sampled subgraphs and run data-parallel GNN on top of them. Similarly, P3~\cite{swapnil2021p3} introduced a dimension-based partitioning with layer-wise model distribution. Sancus~\cite{pen2022sancus} adaptively skips broadcasting to avoid communication in data-parallel GNNs. Recently, a few systems have appeared to tackle temporal GNN models by providing more efficient temporal sampling algorithms for such scenarios (TGL~\cite{zhou2022tgl}, T-GCN~\cite{huan2023tgcn}). DynaGraph~\cite{dynagraph} proposed caching mechanisms for efficient dynamic GNN execution. However, it still tackles only ad-hoc queries and is primarily useful for training workloads.

The GraphTides~\cite{erb2018graphtides} framework was introduced to assess graph-based streaming platforms. Within this framework, streaming graph processing encompasses the capability to manage dynamic graphs (both topology and feature updates) and execute real-time algorithms directly on the graph without external queries. This is achieved while meeting the demands of the streaming domain, such as fault-tolerance, high availability, and scalability. To the best of our understanding, {\Ours} stands out as the premier distributed GNN system adept at handling such workloads. While it might be feasible to partially mimic this use-case on the mentioned systems by periodically inferring the entire graph, the mini-batching in these models leads to uneven loads and postpones the latest inference results by the duration of the batch window at best. Simply put, these systems neither cater to streaming graphs as input nor are they tailored for latency-sensitive applications.



\subsection{Streaming Graph Partitioners}


Partitioning algorithms are widely used in distributed graph processing to enable data-parallelism for load balance and low communication.
Low-latency partitioners are applied at the same time that the graph is being loaded into the cluster~\cite{stanton2012streaming,tsourakakis2014fennel,filippidou2015online}.
This online approach is well-suited to dynamic graphs where offline repartitioning of the entire new graph snapshot is inefficient.
An experimental comparison across different applications with Apache Flink~\cite{abbas2018streaming} shows that data-model-specific techniques (e.g., FENNEL~\cite{tsourakakis2014fennel} for vertex stream, HDRF~\cite{petroni2015hdrf} for edge stream) offer better communication performance while data-model-agnostic methods (e.g., hash) trade off data locality for better balanced workloads.
Despite these studies on various graph algorithms (e.g., shortest paths, PageRank), there is no work that adapts the utility of streaming partitioners for distributed graph learning.
We investigate the effect of streaming partitioning while producing node representations using GNNs on streaming graph data.
\section{Background}
\label{section:background}

This section presents the background needed to cover how {\Ours} provides a fault-tolerant dataflow pipeline, and how it achieves distributed GNN inference and training on streaming graphs.

\subsection{Graph Streams}
\label{subsection:bg-streams}
Real-world graphs are often dynamic in nature, exhibiting changes in their topology as well as node and edge features over time. 
We denote a multi-modal graph as $G = (V, E, X_V, X_E)$ comprising nodes $ v\!\in\!V$, edges $e_{u,v}\!\in\!E \subseteq V\!\times\!V$. 
Additionally, nodes and edges may contain some features. To simplify the presentation, we consider a single feature associated with each node, denoted by $x_v~\forall v\!\in\!V$, and similarly consider edge features $x_e~\forall e\!\in\!E$. We use edge streams to ingest topological data, while node features as feature stream. Nonetheless, the granularity of updates is not limited to this particular paradigm. For example, one can employ vertex streams (streaming nodes along with their local neighborhoods) jointly with  a matching partitioner like Fennel~\cite{tsourakakis2014fennel}. More generally, each streaming event is timestamped and may be a create, delete, or update operation on a graph element (vertex/edge/feature/sub-graph).

\subsection{Distributed Streaming Dataflow}
\label{subsection:bg-flink}
 To build {\Ours}, we leverage Apache Flink\footnote{\url{https://flink.apache.org}}, a stateful stream processing system that started as a research project and is now widely used in the industry for processing data streams. Flink uses a pipeline of data transformation tasks, called operators, to consume input streams and emit output streams. Operators can be parallelized across threads and machines, with each parallel sub-operator instance performing local computations. Fault-tolerance is guaranteed through replayable source operators and intermittent checkpoints to reliable storage. States can be recovered from this storage and used for the re-scaling of partitions. Flink employs a variation of the Chandy-Lamport algorithm~{\cite{chandy1985distributed}} for distributed checkpointing, ensuring that correctness is maintained.

Although Flink pipelines are typically represented as Directed Acyclic Graphs (DAGs), we extend the framework by introducing a novel way to add nested cycles to enable sending back the gradients and performing state replication. This supports fault-tolerant, iterative computations for more complex tasks as part of our solution. We enhance the expressiveness and flexibility of Flink and develop a  set of novel modules to better handle a wider range of streaming use cases. Our end-to-end solution enables efficient and scalable processing of iterative streams, while maintaining fault-tolerance by including in-flight iterative events within checkpoints.

\subsection{Graph Neural Network}
\label{subsection:bg-gnn}
In Graph Neural Networks (GNNs), the learning process often involves the generation of node (or edge) embeddings, which are used to perform downstream tasks such as node classification, link prediction, and similarity queries.
The Message Passing GNN (MPGNN~\cite{gilmer2017neural}) paradigm views the computation task for each node in a GNN layer as a message generation process along incoming edges, followed by an aggregation operation at the receiving node. The node then updates its representation, which is used by the next GNN layer. Below is a common MPGNN formulation where the GNN layer $l$ generates embeddings $x_v$ for layer $l\!+\!1$ at every node $v$:
\begin{align*}
    m_e^{(l+1)} &= \phi(x_u^{(l)},x_v^{(l)}, x_e^{(l)}) \hspace{3em} \forall~(u,v,e) \in N_{in}(v)\\
   a_v^{(l+1)} &= \rho({m_e^{(l+1)}: (u,v,e) \in N_{in}(v)})\\
   x_v^{(l+1)} &= \psi(x_v^{(l)}, a_v^{(l+1)})
\end{align*}
The messages $m_e$ are generated along each incoming edge $e=e_{u,v}$ in the node's in-neighborhood $N_{in}(v)$, as some function of the features ($x_u, x_v, x_e$). All $|N_{in}(v)|$ messages at $v$ are then aggregated into $a_v$, which is used in combination with its old feature $x_v^{(l)}$ to generate the new representation $x_v^{(l+1)}$ for layer $l\!+\!1$.

Adhering to the common message passing GNN (MPGNN) formulation, our approach supports a diverse range of concrete GNN models, such as Graph Convolutional Networks~\cite{kipf2016semi}, GraphSAGE~\cite{hamilton2017inductive}, Graph Attention Networks~\cite{velivckovic2017graph}, and Jumping Knowledge Networks~\cite{xu2018representation}, through the variation of the \textsc{Message} (\(\phi\)), \textsc{Aggregator} (\(\rho\)), and \textsc{Update} (\(\psi\)) components. 
Typically, $\phi$ and $\psi$ are neural networks in themselves, whereas $\rho$ is one of \textit{Concat}, \textit{Sum}, \textit{Mean}, \textit{Min}, \textit{Max} functions. 
A multi-layer GNN is constructed by stacking these computations, and an output layer (implemented as a neural network) is used to generate predictions based on the task.

\subsection{Graph Partitioning}
\label{subsection:bg-partition}
Graph partitioning divides a graph's nodes or edges into disjoint subsets, or partitions, aiming to minimize `cut-edges` or `replicated-nodes`. This NP-Hard problem 
seeks to balance computing loads and reduce network communication. However, most existing algorithms focus on static graphs, making them ill-suited for dynamic, real-time inputs due to lengthy computation times.

In response, {\Ours} involves streaming graph partitioning that compute partitions in real-time as each graph event arrives. We separate logical from physical partitioning for enhanced system scalability. Utilizing HDRF, CLDA~\cite{rad2017clda}, METIS and Random streaming vertex-cut partitioners, we achieve data-parallelism by distributing incoming edge streams. Additionally, {\Ours} presents a feature-granular master-replica synchronization method to manage vertex replication among sub-operators.



\section{Methodology}
\label{section::methodology}

\begin{figure}[t]
  \centering
   \includegraphics[width=\linewidth]{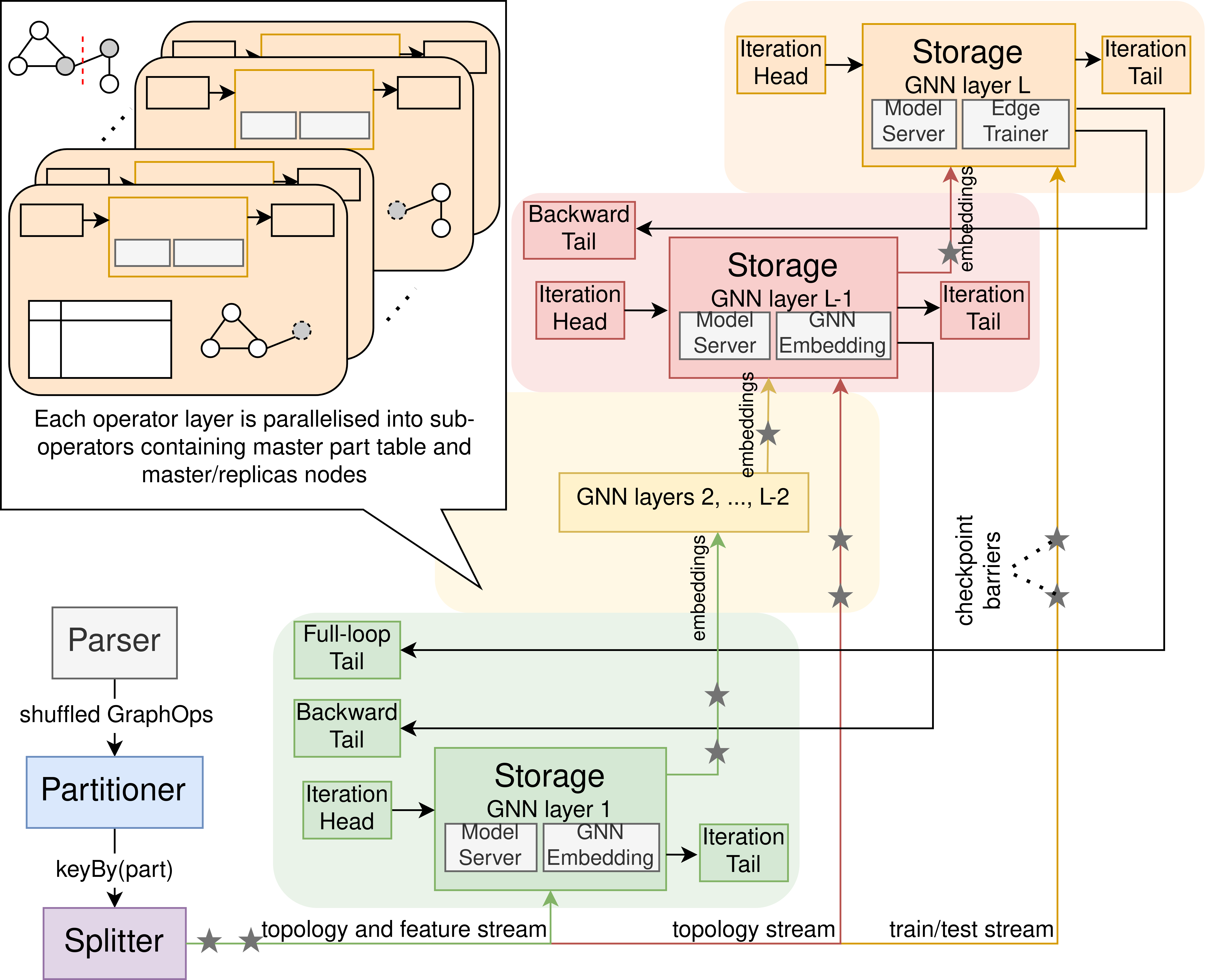}
  \caption{{\Ours} dataflow pipeline}
  \label{fig:gnnpipeline}
  \Description[Full dataflow pipeline]{Pipeline of the operators starting from parser, partitioner, splitter going toward individual gnn storage and processing operators}
\end{figure}

{\Ours} performs asynchronous and incremental inference, enabling the production of up-to-date node representations under streaming graph updates. In cases where re-training is necessary, it performs synchronous backpropagation in a distributed fashion, while avoiding stale states and eliminating bursty resource provisioning. Stale states are a result of outdated information during training, causing asymmetric forward and backward functions. This can occur due to asynchronous updates in graph topology and node embeddings, while backpropagation is taking place, and can negatively impact model accuracy. To be able to tackle this, current systems would maintain a training environment that pulls a static graph snapshot to perform backpropagation, which results in bursty resource provisioning and extensive data migration. Our approach eliminates the need for a separate training environment, reducing the computational burden and improving overall efficiency while providing full-graph, static training guarantees.

\subsection{System Overview}
\label{subsection:system_overview}

{\Ours} involves a dynamic dataflow pipeline, as depicted in Figure~\ref{fig:gnnpipeline}. This pipeline leverages a unified message format that supports create, update, and delete events for edges, vertices, and features. The process begins with a source \texttt{\textbf{Dataset}} parser. While we utilize temporal edge-list files to stream per-edge addition events and per-node features, in practice, any stream source, such as database logs, edge devices, or message queues, can be used. Coupling our system with Flink allows us to seamlessly integrate a broad range of such external data sources. After the events are generated, the streaming \texttt{\textbf{Partitioner}} consumes them, assigning and routing the events to their designated parts. This facilitates data-parallelism for subsequent operators. Before the GNN operators receive them, incoming events pass through a \texttt{\textbf{Splitter}}. This filter improves memory efficiency by ensuring that GNN layers do not process unnecessary events. We categorize events into three classes to cover the majority of inductive use-cases in GNNs:

\begin{itemize}[leftmargin=*]
\item Topology data: Received by all GNN layers, except for the output layer when the task is node-level.
\item Feature data: Only the first GNN layer receives this, as subsequent layers inherit their features from the preceding one.
\item Train \& test data: Solely received by the final output layer for training and reporting real-time metrics.
\end{itemize}

Subsequent GNN layers are distributed into their associated \texttt{\textbf{Graph Storage}} operators, facilitating model-parallelism. Furthermore, this design allows us to control the parallelism of GNN layers independently, thereby better accommodating explosive workloads. These operators incrementally store their graph partition as events arrive. Each sub-operator handles a subset of graph parts and houses the actual GNN layer. In addition, it includes a set of \texttt{\textbf{Plugins}} that monitor local graph updates and execute computations at a granular level, like individual feature updates. This design allows us to devise incremental GNN computations, ensuring that the graph partition and GNN code remain co-located, unlike in sampling-based systems. Such a configuration grants quick access to essential data and provides flexibility for other applications. Moreover, \texttt{Graph Storage} features a tensor manager for rapid native memory management and supports iterations for master-replica synchronization and gradient feedback. We delve deeper into these optimizations in Section \ref{section:system_components}.

In this context, "feature" pertains to graph elements linked to their parent element (e.g., vertices or edges) and holds a value of an arbitrary data type. Embeddings, aggregators, and class labels all qualify as features. Some features might be designated as \textbf{halo}, signifying that they will not be duplicated even if their parent element is replicated.

\subsection{GNN Inference}
\label{subsection:gnn_inference}
\begin{figure}[t]
    \centering
    \begin{subfigure}[h]{\linewidth}
    \centering
        \includegraphics[width=0.25\linewidth]{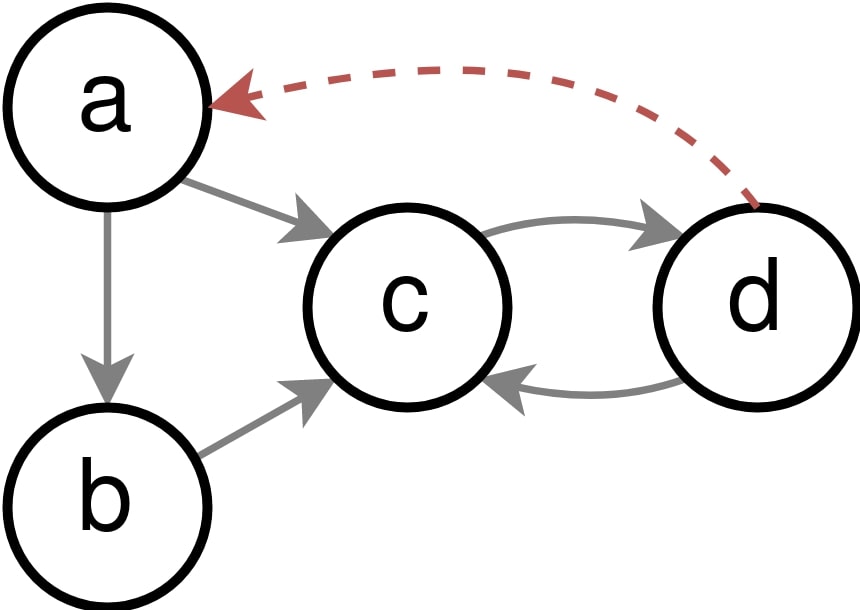
        }
        \caption{Example graph with new edge $d$ to $a$}
        \label{fig:examplegraph}
    \end{subfigure}
    \begin{subfigure}[b]{\linewidth}
    \centering
        \includegraphics[width=0.45\linewidth]{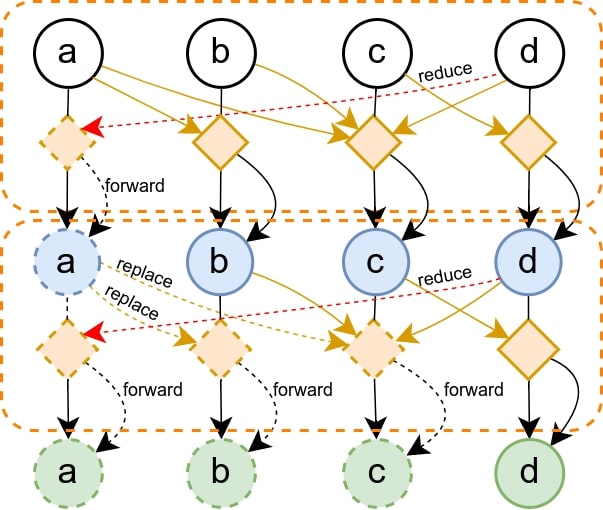}
        \caption{Unrolled computation graph with updates (dotted elements)}
        \label{fig:gnncomputation_updated_edge}   
    \end{subfigure}  
    \caption{GNN Computation Graph}
    \label{fig:computation_graph}
    \Description[Example computation workload]{Description of the 2-layered, streaming GNN computation graph}
\end{figure}

In order to maintain up-to-date node representations in an online fashion, the system must continuously track all influenced nodes $I$ whose representations have become outdated due to new topological or feature update to the graph. This is necessary due to cascading effect of updates during model inference. For a GNN with $L$ layers and an input graph having average in-degree $\delta_{in}$ and average out-degree $\delta_{out}$, an edge addition is expected to influence $|I| = \sum_{l=0}^{L-1}\delta_{out}^{l}$ nodes. Computing the set of influenced nodes is $O(\delta_{out}^{L-1})$ as it requires fetching L-1 out-neighborhood. Updating node representations, in turn, requires retrieving their L-hop in-neighborhood (roots of computation graph) and performing forward pass on them. This in turn means constructing $|I|$ computation graphs each with $\delta_{in}^{L}$ source vertices (i.e., $O(\delta_{in}^{L})$ each). As such, the total cost of supporting a single edge update becomes $O(\delta_{out}^{L-1} + [\sum_{l=0}^{L-1}\delta_{out}^{l}]*\delta_{in}^L)$. Consequently, when dealing with large graphs, performing low-latency inference can quickly become an impractical and difficult task.

In our approach, we avoid explicitly tracking influenced nodes or repeatedly pulling neighborhoods for constructing local computation graphs. Instead, as depicted in Figure~\ref{fig:gnnpipeline}, we treat the chained \texttt{GraphStorage} dataflow as an implicit computation graph, naturally partitioned by breadth (data-parallelism) and depth (model-parallelism), and enable incremental cascades based on changes on graph topology and features. In doing so, {\Ours} achieves a much lower cost $O(\delta_{out}^{L-1})$ of updating influenced node representations with a single edge addition.


\subsubsection{\textbf{Incremental aggregation}}
\label{subsubsection:incremental_aggregation}
In MPGNN (Section~\ref{subsection:bg-gnn}), aggregators summarize the messages that arrive to the node from all of its in-neighbors. These are typically permutation-invariant functions such as sum, mean, and concatenation. We note that, such functions can be incrementally updated by maintaining a relatively small state using exact or probabilistic data structures. To formalize this approach, in {\Ours}, we develop \textsc{\textbf{Aggregators}} as instances of synopsis operations that are cached at each master node to maintain incremental computations. The states of \texttt{Aggregators} are updated by remotely invoking one of the following method interfaces at the master node:\\ 
\-\hspace{2em} $reduce(msg, count=1)$ to add a new message\\
\-\hspace{2em} $replace(msg_{new}, msg_{old})$ to update a message \\
\-\hspace{2em} $remove(msg, count=1)$ to delete a message\\
Our incremental formulation for the \textsc{Aggregator} is customizable, being contingent only on the restrictions of synopsis operators, as they have to be \textit{mergeable, commutative, and invertible}. It can thus support any \textsc{Update} and \textsc{Message} neural network definitions, with their properties having no impact on the incremental functionality.
Moreover, under massive graph updates, we do not require any graph locking mechanisms and allow many cascades to be simultaneously updating the system.
Since the aggregators are permutation-invariant and feature updates follow causally consistent dataflow, the incremental model is \textbf{eventually consistent}.

Therefore, the tracking of influenced nodes and the retrieval of node neighborhoods occur automatically, by following the dataflow between operators. These are triggered by external updates at each layer of the GNN.

Most temporal GNN architectures make use of memory modules which are new embeddings that 
do not conflict with the above essential properties of the aggregators. 
This allows our model to be adaptable to temporal GNNs without compromising the integrity of its core functions. 
Integrating nonlinear recurrent units like LSTMs and GRUs, which diverge from the traditional aggregator paradigm, presents a more complex challenge. These units, by design, necessitate the retention of an extensive message history for each node. 
A permutation-invariant version of this approach would enable us to streamline the model by retaining only the most recent hidden state for each aggregator, aligning seamlessly with the operational protocols of our described {\Ours} interfaces. For instance, the process for replacing a message could be elegantly bifurcated into two distinct phases: initially, the model would apply the sign inverse of the old message to effectively nullify its impact, followed by the computation of the new message.

\subsubsection{\textbf{Streaming forward pass}}
\label{subsubsection:streamingforwardpass}

This method describes pure streaming approach for inference, where the next layer representations are immediately updated by cascading through the computation graph as described earlier. The arriving edges cause \textsc{Messages} to be sent to destination aggregators. As the \texttt{Aggregators} receive updates, the algorithm generates up-to-date representations through \texttt{Update} function and forwards it to the next \texttt{GraphStorage} along the chain. Note that, we employ vertex-cut partitioning in our system, therefore some vertices are replicated and the corresponding \textsc{Aggregator} resides only with the master vertex (and not with replicas). This distributes edge-based computations across machines, achieving greater load balance. 

To illustrate, consider Figure~\ref{fig:gnncomputation_updated_edge}, which unfolds the computation graph generated, based on MPGNN formulation, given the example graph in \autoref{fig:examplegraph}. The dotted lines represent edge addition event and the cascading computations taking place in the computation graph respectively. The illustrated computation graph represents layers sequentially from top to bottom, i.e., node representations are seen for input and two GNN layers. \textsc{Aggregators} are diamonds. 
Orange arrows directed towards \textsc{Aggregators} depict \textsc{Messages}, and black arrows towards nodes are \textsc{Updates}. For example, the new edge from node $d$ to $a$ (\autoref{fig:examplegraph}) affects the representations for vertices $a$, $b$, and $c$. This can be traced by looking at affected leaf nodes (dashed borders) in computation graph. 

Algorithm~\ref{streamingInference} illustrates the pseudo-code for streaming inference for a simplified MPGNN-style model (GraphSAGE) which is uni-modal and does not contain edge features.
$createAggregator()$ creates and attaches \texttt{Aggregator} to its vertex in all gnn layers. 
$msgReady()$ and $updReady()$ check if all the data dependencies are in place for \textsc{Message} and \textsc{Update} functions to be computed.
$reduce()$ and $replace()$ functions send \texttt{Remote Method Invocation} messages to the corresponding \textsc{Aggregators} at the master. 
$forward()$ function computes the next layer representation $x_u^{l+1}$ and sends it to the subsequent operator as vertex feature update.

\begin{algorithm}[h]
\caption{Streaming Inference}\label{streamingInference}
\begin{algorithmic}[0]
\Require node $u$ with feature $u.f$ and \textsc{Aggregator} $u.agg$, out-neighborhood $(u,v,e) \in N_{out}(u)$, functions \textsc{Message} (\(\phi\)), \textsc{Update} (\(\psi\))
\Function{$addElement$}{$u$}
\IfThen{$u.state()==MASTER$}{$createAggregator(u)$}
\EndFunction
\Function{$addElement$}{$e$}
\IfThen{$msgReady(e)$}{$v.agg.reduce(\phi(e))$}
\EndFunction
\Function{$addElement$}{$u.f$}
\IfThen{$updReady(u)$}{$forward(\psi(u.f, u.agg))$}
\ForAllDo{$N_{out}(u)$}{$v.agg.reduce(\phi(e))$}
\EndFunction
\Function{$updateElement$}{$u.f^{new}, u.f^{old})$}
\IfThen{$updReady(u)$}{$forward(\psi(u.f^{new}, u.agg))$}
\ForAllDo{$N_{out}(u)$}{$v.agg.replace(\phi(e^{new}), \phi(e^{old}))$}
\EndFunction
\Function{$updateElement$}{$u.agg^{new}, u.agg^{old}$}
\IfThen{$updReady(u)$}{$forward(\psi(u.x, u.agg))$}
\EndFunction
\end{algorithmic}
\end{algorithm}

The streaming setting can cause three possible bottlenecks: 
(i) Neighborhood explosion in GNNs
can cause \texttt{Graph Storage} operators at deeper layers to receive higher workload, increased by a factor of $\delta_{out}$ with every additional layer.
(ii) Vertices with high centrality scores (hub vertices, especially in power-law graphs) emit new features more frequently, hence can overwhelm the subsequent sub-operators. 
(iii) Changing external workload patterns (e.g., because of seasonality, real-world events) can concentrate graph updates in a narrow region of its topology.

\subsubsection{\textbf{Explosion Factor}}
\label{subsubsection:explosion_factor}

Since our GNN layers are fully decoupled and our graph is only logically partitioned, we can vary the \textit{parallelisms} of \texttt{Graph Storage} operators independently. Hence, to tackle GNNs' neighborhood explosion challenge, we introduce a new system hyper-parameter called `\textbf{\emph{explosion factor}}' ($\lambda$). This enables us to vary the parallelism $p_i$ of each individual \texttt{Graph Storage} operator, i.e., the number of sub-operators that perform the same task in a data-parallel manner.
Namely, given an initial parallelism $p$ and $L$ layers of the GNN, we assign the actual parallelism for each \texttt{Storage} operator (layer)
as $p_{i}=p*\lambda^{i-1}$ for $i\in[1, \dots, L]$. 
This parameter must be selected considering the frequency of training, as even though the forward pass is always benefited by higher $\lambda$, because neighborhood explosion has reverse effect on layer-wise workload during backward pass for training.

\subsubsection{\textbf{Windowed forward pass}}
\label{subsubsection::windowed_forward_pass}
We propose \textbf{intra-layer} and \textbf{inter-layer} windowing to mitigate neighborhood explosion, data skews, and master node imbalances. Unlike the streaming algorithm (Algorithm \ref{streamingInference}), which executes the $forward$ and $reduce$ functions immediately, the windowing approaches introduce a time-based window that delays their execution.

Intra-layer windowing (which delays the $forward$ functions for each vertex) is especially beneficial for hub-vertices. It allows us to send a single, most up-to-date update for a batch of $forward$ requests. By doing this, we can optimize network usage and reduce cascades in the subsequent layer, thereby minimizing effects of neighborhood explosion.

Although our system utilizes vertex-cut partitioners to balance per-edge computations, the presence of nodes that receive updates only at master nodes can introduce additional skews. Thus, inter-layer windowing delays the emission of reduce messages for each destination node. It batches the corresponding edges, calculates a local, partial aggregation for each destination node, and emits a single reduce message summarizing the batched edges.

Algorithm~\ref{abstract_windowed_inference} provides the pseudo-code for the abstract windowed forward pass. Depending on the $intraLayerWindow$ and $interLayerWindow$ functions, we propose three windowing algorithms. We employ timers to manage windowing with a 10ms coalescing interval, ensuring that timer threads are not overwhelmed.

In the \textbf{Tumbling Windowed} approach, each $forward$ node and $reduce$ destination are allocated a window of a specific duration. This naturally enables the batching of intra and inter-layer computations based on the frequency of the corresponding cascades during its interval. By adjusting the window interval, we can effectively control the latency overheads of this method.

Shifts in dynamic patterns can cause some edges to become highly active for a specific duration. For instance, a concert might lead to a sudden surge in merchandise sales. Such phenomena can introduce workload skews in sub-operators, even when using tumbling windowing. To address this, we propose \textbf{Session Windowing}. In this approach, the aforementioned functions are evicted after a certain, fixed period of inactivity. This algorithm is similar to the tumbling one, with the primary distinction being that adding a vertex already in a window further postpones its eviction time.

Nodes in the real world can exhibit varied and evolving frequencies. Therefore, the inactivity duration considered a "session" for a specific node is dynamic. To account for this, we introduce \textbf{Adaptive Session Windowing}, where session intervals are determined based on the windowed exponential mean of past frequencies. To enable low-storage computations of the windowed exponential mean, we designed a thread-safe CountMinSketch that is periodically averaged.

\begin{algorithm}[h]
\caption{Windowed Inference}\label{abstract_windowed_inference}
\begin{algorithmic}[0]
\Require node $u$ with feature $x_u$ and \textsc{Aggregator} $agg$, out-neighborhood $(u,v,e) \in N_{out}(u)$, $forwardBatch$ holding delayed forward vertices, $reduceBatch$ holding delayed reduce edges per dest vertex, functions \textsc{Message} (\(\phi\)), \textsc{Update} (\(\psi\))\\
\Comment{Other functions same as streaming algorithm}
\Function{$addElement$}{$e$}
\If{$msgReady(e)$}
    \State $e.delete()$
    \State $intraLayerWindow(e)$
\EndIf
\EndFunction
\Function{$forward$}{$vertex$}
    \State $interLayerWindow(vertex)$
\EndFunction
\Function{$evictForward$}{$timestamp$}
    \State $vertices {\gets} forwardBatch.lessThan(timestamp)$
    \State $updates {\gets} \psi(vertices.f, vertices.agg)$
    \For {$i \in [0, vertices.len)$}
        \State $send(updates[i], vertices[i].master)$
    \EndFor    
\EndFunction
\Function{$evictReduce$}{$timestamp$}
    \State $edges {\gets} reduceBatch.lessThan(timestamp)$
    \State $edges.create()$
    \State $srcMessages {\gets} \phi(edges.src.unique.f)$
    \State $reduceMsgs {\gets} scatterAggregate(srcMessages, edges)$
    \For {$ (dest, count, index) \in edges.groupby("dest")$}
        \State $dest.agg.reduce(reduceMsgs[index], count)$
    \EndFor
\EndFunction
\Function {$onTimer$}{$timestamp$}
    \State $evictReduce()$
    \State $evictForward()$
\EndFunction
\end{algorithmic}
\end{algorithm}

\subsection{GNN Training}
\label{subsection:gnn_training}
Aside from causing load imbalance, external workloads occasionally change the distribution of "true" representations. This phenomenon is more commonly referred to as "concept drift." To stay abreast of these changes, model re-training becomes necessary. In this section, we describe our approach in {\Ours} that allows for this re-training on the same cluster while preventing staleness.

The distributed training must be coordinated to ensure there are no inconsistencies when performing the backward pass during streaming graph updates. Gradients sent back through the computation graph will become invalid if the topology changes during this time. Furthermore, after the training is concluded, intermediate node embeddings and aggregators need to be recalculated to mirror the updated model.

To address this, we introduce a specialized, fault-tolerant \texttt{\textbf{Training Coordinator}} process within the job manager. This process oversees the entire GNN training life-cycle, which includes initiating and terminating the distributed training loop, computing epochs and batch sizes, and averting staleness.

In GNN training, we append an output \texttt{GraphStorage} operator following the final embedding layer. This operator captures the final node representation, true labels, and a loss function ($\mathcal{L}$). During job definition, the loss function is seamlessly integrated into the designated trainer \texttt{Plugin}. Concurrently, the true labels and node representations are introduced as streams, in line with the approach detailed in Section~\ref{subsection:system_overview}.

\begin{figure}[t]
    \centering
        \includegraphics[width=0.75\linewidth]{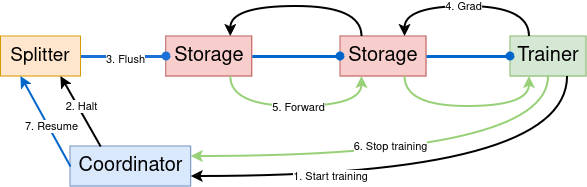}
        \caption{Distributed Training Overview}
        \label{fig:distributed_training}   
    \Description[Overview of distributed training]{High-level distributed training loop coordinated across all GNN operators}
\end{figure}

An overview of the sequence of events in distributed training is presented in Figure~\ref{fig:distributed_training}. The incoming stream is paused at the \texttt{Splitter}, allowing all messages to be processed through the pipeline. Once paused, the requisite epoch and batch counts are determined based on the volume of training data available. This is followed by distributed backpropagation, synchronization of model parameters, and a full-batch forward pass. Upon completion of the training, {\Ours} reactivates the \texttt{Splitter} and resumes its standard operation using the refreshed model. Subsequent portions of this section delve deeper into each stage.


\subsubsection{\textbf{Coordination of distributed training}}
\label{subsubsection:coordtrain}
The distributed sub-operators of the final layer need to coordinate to determine the start and end times for their training. To initiate this process, we employ a majority-voting mechanism, wherein the coordinator begins the training loop once more than half of the available output sub-operators signal the \texttt{StartTraining} command.

The decision to start training, made by the output sub-operators, could be either periodic (e.g., in {\Ours}, it is triggered when a pre-defined batch size is reached in the final \texttt{Storage} operator) or adaptive (e.g., based on test performance).

Upon entering the training mode, the coordinator halts the \texttt{Splitter} from consuming external updates. Thanks to its streaming design, such stoppage gradually cascades back to the stream source (\texttt{Dataset}), preventing memory bloating issues for prior operators. It is important to note that iterative messages still flow freely during the training phase.

The asynchronous pipeline implies that there might still be in-flight (unprocessed) events flowing through the pipeline. To flush out these remaining messages, the coordinator employs a termination detection algorithm as described in Section~\ref{section:system_components}. Correctly flushing the pipeline, combined with the halt of external updates, ensures that staleness issues do not arise during backpropagation.

Lastly, before initiating the training loops, the output sub-operators share their training data sizes, aggregate them, and then determine a batch size.  We choose epochs to be static during the definition of the pipeline.

\subsubsection{\textbf{Distributed backpropagation}}
\label{subsubsection:dist_backprop}
Once the computation graph is frozen, backpropagation begins in each logical partition of the final (output) layer:
\begin{enumerate}[leftmargin=*]
\item Fetch the prediction layer inputs for each train label from the current batch (node embeddings for node-based tasks or source and destination node embeddings for edge-based tasks).
\item Perform predictions and evaluate the loss function $\mathcal{L}$.
\item Run the local backpropagation algorithm to generate model gradients as well as embedding gradients $\partial{\mathcal{L}}/\partial{x_v^{(L)}}$.
\item Send $\partial{\mathcal{L}}/\partial{x_v^{(L)}}$ back to each corresponding master vertex in the previous layer.
\end{enumerate}

Backprop is triggered (via broadcast instructions) in the previous layer operator only after all logical parts of the current sub-operator complete their job. Each logical part accumulates (per-vertex) the $\partial{\mathcal{L}}/\partial{x_v^{(l+1)}}$ for the layer that gets sent back to it. Once a broadcast instruction is received by a layer from all sub-operators, no more embedding gradients are expected. That layer is then free to perform its backpropagation in two phases. The first phase, within which computations can proceed asynchronously, is as follows:
\begin{enumerate}[leftmargin=*]
\item Compute ${x_v^{(l+1)}}$ for each accumulated vertex in the logical part.
\item Perform local backprop (i.e., Jacobian-vector product) using stored \textsc{Aggregator} state $a_v^{(l+1)}$, vertex feature $x_v^{(l)}$, and accumulated gradients to get $\partial{\mathcal{L}}/\partial{a_v^{(l+1)}}$ and $\partial{\mathcal{L}}/\partial{x_v^{(l)}}$.
\item Send $\partial{\mathcal{L}}/\partial{x_v^{(l)}}$ back to each corresponding master vertex.
\item Send $\partial{\mathcal{L}}/\partial{a_v^{(l+1)}}$ to all replicas of the given vertex, along with $a_v^{(l+1)}$ for gradient computation (since replicas do not have them locally).
\end{enumerate}

The second phase then proceeds (also asynchronously) as follows:
\begin{enumerate}[leftmargin=*]
\item Once all the \textsc{Aggregators} are received, compute the \textsc{Messages} $m_e^{(l+1)}$ from the locally available in-edges.
\item Compute $\partial{\mathcal{L}}/\partial{m_e^{(l+1)}}: (u,v,e) \in N_{in}(v)$ from $\partial{\mathcal{L}}/\partial{a_v^{(l+1)}}$ and local \textsc{Messages} at the \textsc{Aggregator}.
\item Continue the backpropagation by calculating the $\partial{\mathcal{L}}/\partial{x_u^{(l)}}$ and $\partial{\mathcal{L}}/\partial{x_v^{(l)}}$ using the above message gradients collected.
\item Send gradients back to the corresponding master vertex.
\end{enumerate}

Once the vertex gradients are received the process repeats until the first GNN layers is reached. We make use of the cached \textsc{Aggregator} and vertex feature states that had been calculated and stored during the last forward pass (before training was triggered). Such caching minimizes redundant communication and computations during training. Moreover, our incremental \textsc{Aggregators} calculate gradients using only locally available data and their stored state. Synchronous training phases also allow the use of \textit{vectorization} to perform matrix operations efficiently in bulk.
\subsubsection{\textbf{Model synchronization and forward pass}}
\label{subsubsection:model_sync}
In the first layer, instead of sending back gradients, the system starts the model update and forward pass cycle to recompute up-to-date embedding representations and \textsc{Aggregator} states. This procedure is similar to the streaming forward pass. However, since our graph is now static (external updates are halted due to buffering the incoming graph stream), we introduce several optimizations with layer-by-layer computations in three synchronous phases:

\paragraph{Phase 1 (Model Update)}
Since our model is distributed across sub-operators, the gradients and model parameters must be synchronized after training. Each distributed model runs its local optimizer (e.g., SGD, Adam, Adamax) to update its model parameters. Vertex embeddings are also updated if trainable (i.e., if $x_{v}^0$ are not received as input), which triggers their replica synchronization. Once completed, each sub-operator broadcasts its local parameters to other sub-operators, which compute the mean of the received values. Algorithm~\ref{alg:modelupdate} provides pseudo-code for this process.
\begin{algorithm}
\caption{Model update}
\label{alg:modelupdate}
\begin{algorithmic}[0]
\Function{updateModel}{}
\State $W_i^+ = optimizer(W_i,\Delta{W_i})$ \Comment{In each logical part $i=1, \ldots, P$}
\State $broadcast(W_i^+)$
\State $W^+ \gets collect()$
\State $W_{i} = \dfrac{1}{P}\sum_{j=1}^{P} W_j^+ $
\Comment{In each logical part $i=1, \ldots, P$}
\EndFunction
\end{algorithmic}
\end{algorithm}
\paragraph{Phase 2 (Aggregate)}
Once the model is updated, we can safely continue re-building our computation graph. This involves computing \textsc{Messages} and performing $reduce()$ at each \textsc{Aggregator}, as done in the inference case. However, since the graph is now static, separate synchronous phases for aggregation and update can avoid redundant \textsc{Update} messages. In this phase, we only update the \textsc{Aggregators}, without producing next layer embeddings.

The aggregate phase starts with each master vertex resetting its \textsc{Aggregator} state (usually to a zeros tensor). We perform a \textbf{$batchReduce()$} on all locally available in-edges, and send only the resulting $reduce()$ message to the master \textsc{Aggregator}. As such, the number of $reduce()$ messages sent per vertex is only proportional to its replica counts, not in-degree.

\paragraph{Phase 3 (Update)}
Once the model is guaranteed to be updated and all \textsc{Messages} reduced, all the local master vertex embeddings must be updated. The \textsc{Update} is invoked for the vertices and sent to the next operator, from where a new synchronous cycle begins. This continues up to the final layer. At this point, the \texttt{StopTraining} instruction is activated, prompting the \texttt{Splitter} to resume and transitioning the system back to inference mode.

\subsection{Streaming Graph Partitioning}
\label{subsection:partitioning}
The above inference and training methodologies allow for incremental updates and caching within logical parts alongside master/replica synchronization steps. We now present our partitioning scheme to support the distributed execution of our hybrid-parallel pipeline. 
In particular, we discuss how we optimize the latency of a streaming partitioner operator for our purposes, and how this operator is used to assign as well as re-scale parts.

When distributing the workload, 
the \texttt{Partitioner} operator identifies the correct destination sub-operators by assigning \texttt{part} numbers to the incoming stream data. 
\subsubsection{\textbf{Distributed partitioner logic}}
\label{subsubsection:distribted_partitioner_logic}
We build {\Ours} to utilize any streaming partitioning algorithm within its \texttt{Partitioner} operator. In our implementation, we utilize HDRF, CLDA and Random vertex-cut streaming partitioners. Distributing those requires a shared-memory model for storing partial degree and partition tables, which is not supported by Flink. Without it, a single thread needs to be allocated to the partitioner, which causes a significant bottleneck when scaling up the system. Hence, we develop a novel \texttt{Partitioner} operator to support correct, concurrent thread distribution for streaming partitioners. 
It distributes the main partitioning logic among arbitrary number of threads while having synchronized access to the output channel. 
The latter is necessary to avoid corrupt data during network transfer, as output channels consume data in smaller units than the graph data being streamed. A vertex-locking mechanism is also developed for correctness, where edges with common vertices are assigned to their logical parts one at a time. 



Our \texttt{Partitioner} maintains a master part table where
we can store the first \texttt{part} that an element is assigned to. 
Because vertex-cut partitioning is used, it results in replicated vertices that are assigned to parts different from the first master vertex.
The master part table enables replicas to sync and communicate with their masters from their respective \texttt{Graph Storage} operators. 
Algorithm \ref{streamingPartitioner} describes the pseudo-code for our streaming partitioner.
\begin{algorithm}
\caption{Streaming Partitioner}\label{streamingPartitioner}
\begin{algorithmic}[0]
\Require $state, master\_table, num\_partitions, operator$
\State $part \gets \textit{assignPart(state, master\_table, num\_partitions, operator)} $
\If{$master\_table(operator.element) = \emptyset $}
\State $master\_table(operator.element).insert(part)$
\EndIf
\State $assignMaster(operator.element, master\_table)$
\State $operator.part \gets part$
\State \Return $operator$
\end{algorithmic}
\end{algorithm}

\subsubsection{\textbf{Re-scaling logical parts}}
\label{subsubsection:re_scaling_parts}
Assigning physical partitions alone does not allow flexible re-scaling of \texttt{Graph Storage} operators (e.g., if the number of physical partitions changes due to failure), nor does it support different parallelisms across the chained \texttt{Graph Storage} operators (e.g., to better cope with the exponential load induced by neighborhood explosion).
To tackle this issue, we define the total number of available parts ($num\_partitions$) to be the same as the maximum possible parallelism of the system ($max\_parallelism$), while actually partitioning the 
graph events using \sloppy $keyBy(operator.part)$. In other words, the streaming \texttt{Partitioner} assigns only \textit{logical parts} while the \textit{physical part} is computed using a hash of the assigned logical part. 
As a consequence, multiple logical \texttt{parts} may map to the same sub-operator.
Flink treats the logical parts (keys) in complete isolation; each part maintains its own context (state tables, timers, etc). 

Operators store data (state) in two ways: \texttt{\textbf{Operator State}} stores data for a given sub-operator, which can be accessed by all elements arriving at sub-operator, while \texttt{\textbf{Keyed State}} stores data at the granularity of a unique key, 
and each arriving element can only access data that is assigned to its particular key.
Upon re-scaling {\Ours}, 
\texttt{Operator State} is either randomly redistributed to new sub-operators, or broadcast (in entirety) to all remaining sub-operators to then perform recovery logic. \texttt{Keyed State}, however, is distributed to the new sub-operator containing that key. 
Our flexible mapping of keys to sub-operators allows for re-scaling of the physical partitions based on availability, and a fixed hash function (for logical to physical parts) guarantees fault-tolerant recovery.
Hence, we are able to delegate the fault tolerance logic to Flink and ensure state redistribution and correct operation even under variable parallelisms.


When operator's \textit{parallelism} gets closer to $max\_parallelism$, some sub-operators may remain constantly idle due to never being assigned with any logical parts.
To tackle this, instead of using Flink's default \textit{Murmurhash}\footnote{\url{https://sites.google.com/site/murmurhash/}} algorithm on top of the key's \textit{hashCode}\footnote{Java Object method which generates an integer hash depending on implementation} to compute physical parts, 
we develop Algorithm~\ref{logical_key_assignment_flink}.
Each operator is assigned at least one key, and logical parts are evenly distributed to operators depending on the current \textit{parallelism}. 


\begin{algorithm}
\caption{Compute physical part from logical}
\label{logical_key_assignment_flink}
\begin{algorithmic}[0]
\Require $logical\_part, parallelism, max\_parallelism $
\State $key\_group \gets \textit{logical\_part \% max\_parallelism} $
\State $physical\_part \gets \textit{key\_group * parallelism / max\_parallelism} $
\State \Return $physical\_part$
\end{algorithmic}
\end{algorithm}

\section{System Components and Optimizations}
\label{section:system_components}
This section describes the system-level functionalities and optimizations we introduce within {\Ours} to further facilitate streaming GNN pipeline. 

\subsection{Communication}
\label{subsection:comm}

To enhance communication efficiency within {\Ours}, we introduce custom serializers for frequently used data types, including vertex, edge, and remote method invocation. Additionally, for tensor serialization, we employ compression techniques. We also introduce a $selectiveBroadcast$ primitive, which enables broadcasting an event to specific portions of the graph. For instance, replicating a vertex from a master to its replicas. This method circumvents the repeated serialization typically found in P2P communication, thus conserving computational resources.

To ensure fault-tolerant, iterative communication in {\Ours}, we use in-memory, SPSC, array-based queues. We apply a unified \texttt{\textbf{IterationHead}} logic to wrap operators with terminal feedback edges, which allows for thread-safe event consumption without racing with external ones. Events directed to specific head-operators are collected in separate \texttt{\textbf{IterationTail}} operators, which are co-located with their respective heads to maintain their queues. Our iteration model can handle nested, multi-layered iterations. Fault-tolerance is achieved by stopping queue consumption from heads and resuming it from tails after checkpointing in-queue messages.

\subsection{Storage}
\label{subsection:storage}
{\Ours} involves a custom in-memory storage backend which takes advantage of unboxed data structures. The backend uses two adjacency lists (one for in-edges and one for out-edges) to store edges. A \textbf{task-manager-local storage} option is provided to avoid duplicating data in the cluster. This serves for storing vertex master tables and any other global data structures like CountMinSketch for Adaptive Windowing plugin.

Dealing with tensor garbage collection at large scales could result in excessive memory consumption and potential heap overflows, primarily because their memory is allocated outside the JVM. To counteract this challenge, we've established a per-thread \textbf{tensor cleaning} module grounded on counter-based caching. As each sub-operator processes a unit-event, newly spawned tensors are temporarily stored in local cleaners. Each cleaner updates its counter based on the volume of new tensors and deallocates those surpassing a specific age threshold. This mechanism ensures a restricted count of active tensors, leading to optimized memory utilization.

\subsection{Termination Detection}
\label{subsection:termination}
Introducing nested iterations in {\Ours} requires a valid distributed termination detection algorithm. For that, similar to \texttt{TrainingCoordinator} (Section~\ref{subsection:gnn_training}), we develop a \texttt{\textbf{TerminationCoordinator}}. It periodically collects termination states from all iteration ``heads'' and proceeds with regular dataflow termination once all are ready to be terminated. The ``heads'' are ready to be terminated if they have not received events since the last collection and if they have not got scheduled timers. Last condition is necessary to avoid staleness in window-based inference methods (Section~\ref{subsubsection::windowed_forward_pass}). This algorithm is also used to flush the pipeline in the case of GNN training (Section~\ref{subsection:gnn_training}).

\section{Performance Evaluation}
\label{experiments}
\begin{figure*}[ht]
    \centering
    \captionsetup{font=small} 
    \begin{subfigure}[b]{0.5\textwidth}
        \centering
        \includegraphics[width=\textwidth]{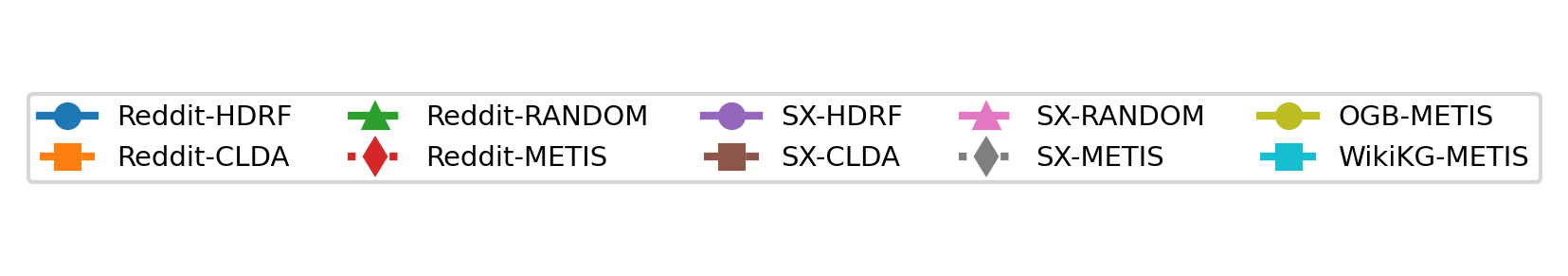}
    \end{subfigure}
    \begin{minipage}{\textwidth}
        \centering
        \begin{subfigure}[t]{0.49\textwidth}
            \includegraphics[width=\textwidth]{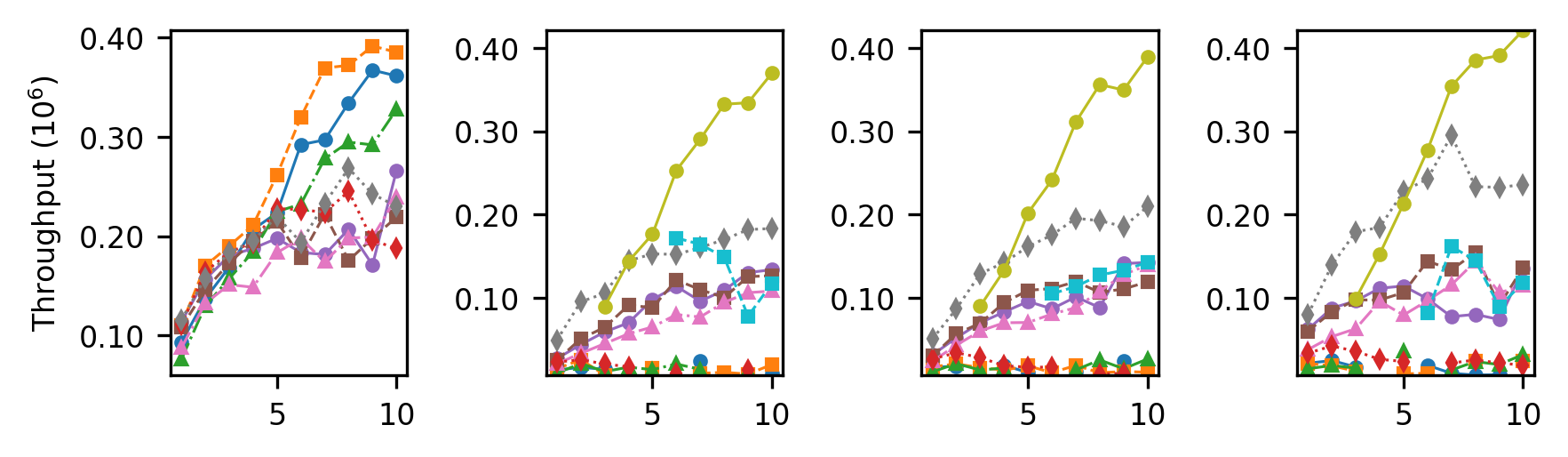}
            \caption{Mean throughput}
            \captionsetup{font=scriptsize}
            \label{fig:sx_superuser_throughput_mean}
        \end{subfigure}
        \hfill
        \begin{subfigure}[t]{0.49\textwidth}
            \includegraphics[width=\textwidth]{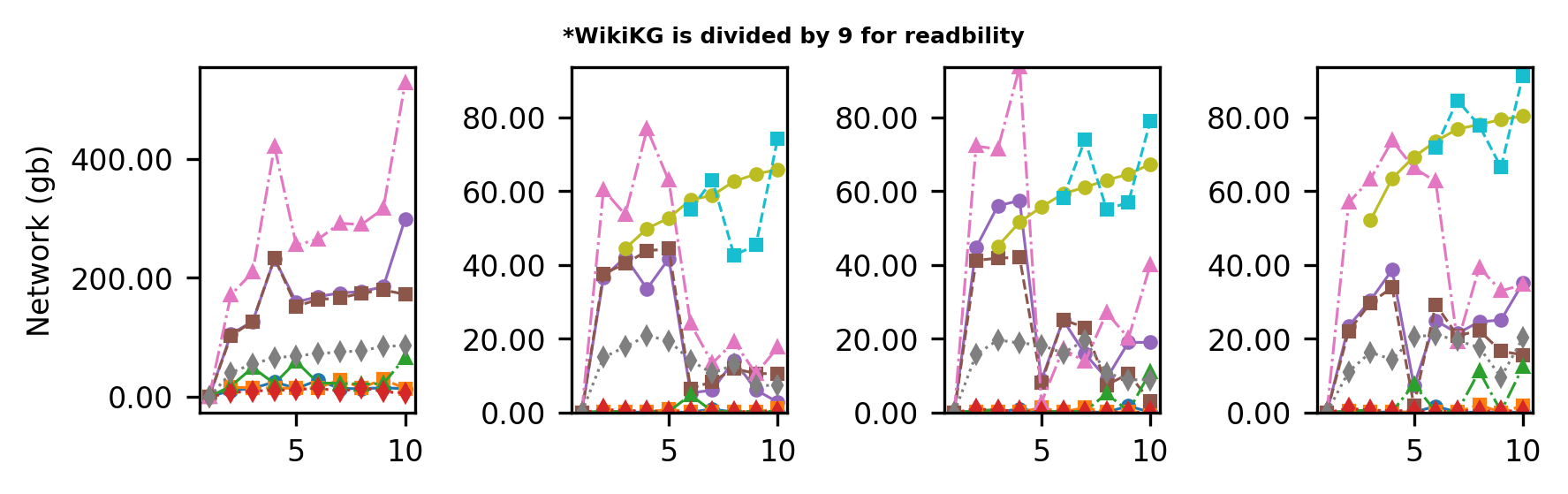}
            \caption{Network volume}
            \captionsetup{font=scriptsize}
            \label{fig:sx_superuser_comm_volume}
        \end{subfigure}
    \end{minipage}
    \begin{minipage}{\textwidth}
        \centering
        \begin{subfigure}[t]{0.49\textwidth}
            \includegraphics[width=\textwidth]{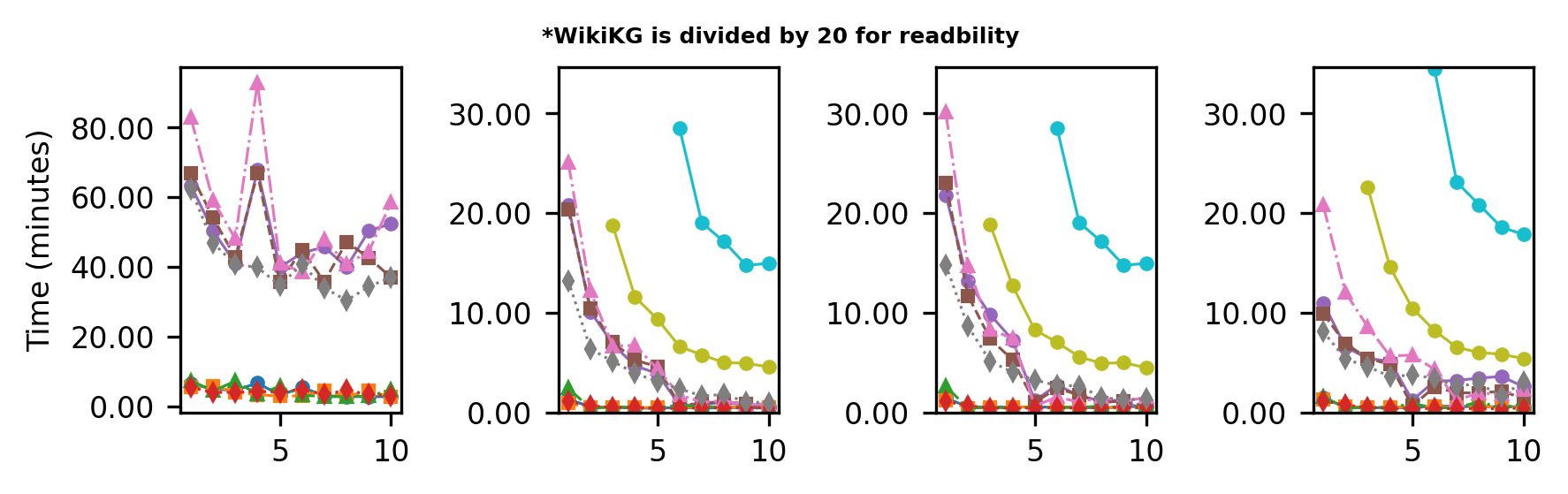}
            \caption{Runtime}
            \captionsetup{font=scriptsize}
            \label{fig:sx_superuser_runtime}
        \end{subfigure}
        \hfill
        \begin{subfigure}[t]{0.49\textwidth}
            \includegraphics[width=\textwidth]{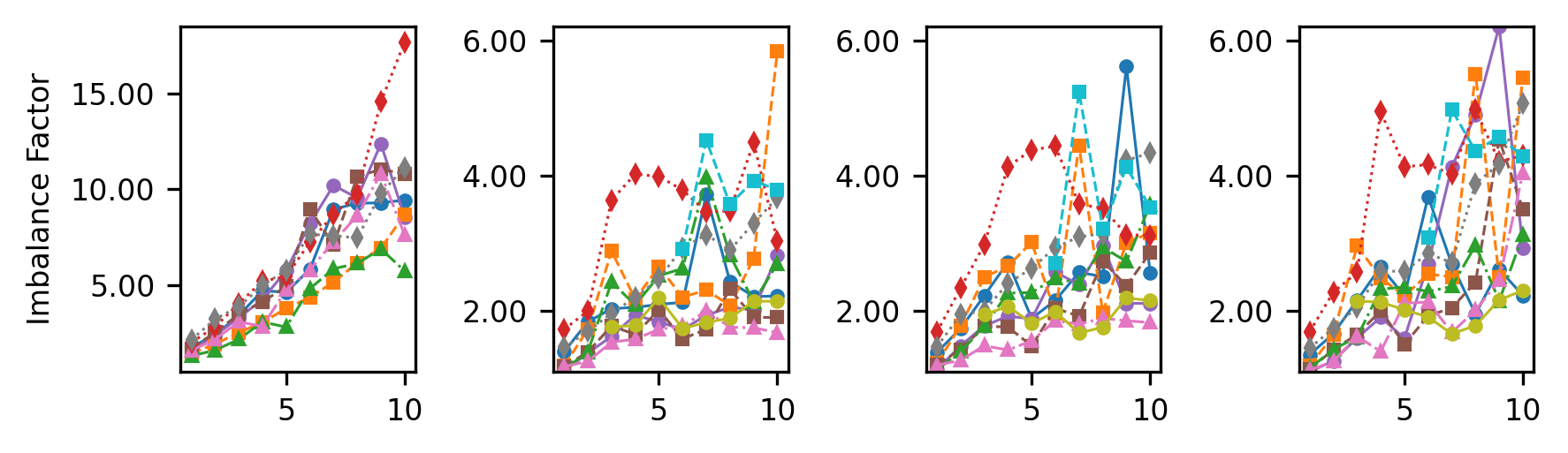}
            \caption{Imbalance Factor}
            \captionsetup{font=scriptsize}
            \label{fig:sx_superuser_imbalance_factor}
        \end{subfigure}
    \end{minipage}
    \caption{Scalability of inference algorithms: Streaming, Session, Sliding, and Adaptive.}
    \captionsetup{font=small}
    \label{fig:sx_superuser}
    \Description[Scalability of \Ours]{Experiments results for the scalability of various inference approach from streaming GNN}
\end{figure*}

\noindent\textbf{\textit{Datasets.}}
To test the performance of {\Ours} on streaming graphs, we use five datasets: \texttt{sx-superuser}~\cite{paranjape2017motifs}, \texttt{reddit-hyperlink}~\cite{kumar2018community}, \texttt{stackoverflow}~\cite{paranjape2017motifs}, \texttt{ogb-products}~\cite{chiang@2019clustergcn} and \texttt{wikikg90Mv2}~\cite{hu2020ogb}.
The data is treated as an incoming stream of edge addition and feature update events to the graph, ordered by the edge timestamps. \texttt{sx-superuser} is temporal network of user interactions on a stack exchange website. It contains 1.4M edges and 200k nodes. \texttt{reddit-hyperlink} dataset contains an edge-list of directed subreddit mentions derived from the Reddit Social Network, with 286K edges and 36K nodes. \texttt{stackoverflow} dataset has question-answers and comments from the StackOverflow social network, with 63.5M edges and 2.6M nodes. \texttt{ogb-products} dataset is a co-purchasing network from Amazon with 62M edges and 2.6M nodes. \texttt{wikikg90Mv2} is a knowledge-graph dataset spanning 601M edges and 91M nodes.
Edge deletion events are also supported in {\Ours} but are not present in the datasets evaluated.

\begin{figure*}[ht]
   
    \begin{minipage}{0.69\textwidth}
        \begin{subfigure}[ht]{\textwidth}
            \centering
             \includegraphics[width=0.6\textwidth]{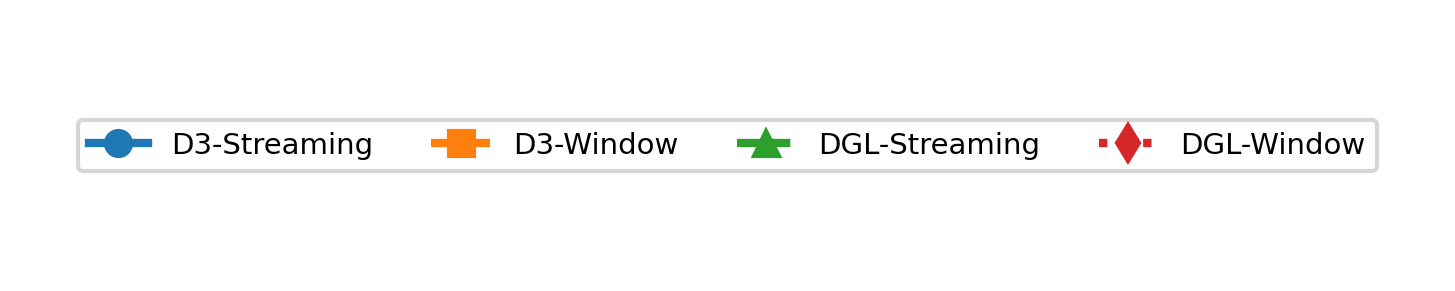}
        \end{subfigure}
        \centering
        \begin{subfigure}{0.24\textwidth}
            \includegraphics[width=\textwidth]{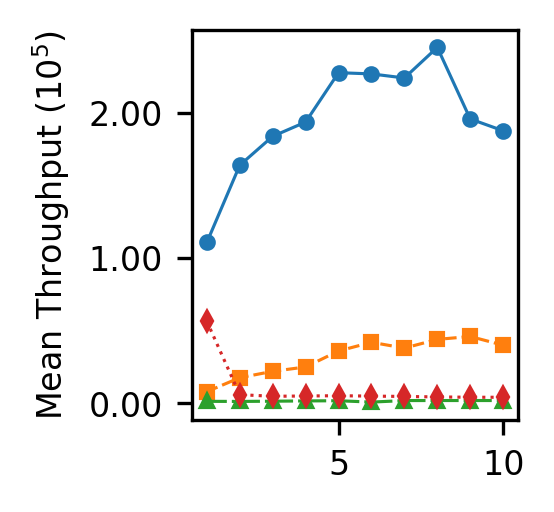}
             \captionsetup{font=scriptsize}
            \caption{Mean throughput}
            \label{fig:reddit_hyperlink_throughput_mean}
        \end{subfigure}
        \begin{subfigure}{0.24\textwidth}
            \includegraphics[width=\textwidth]{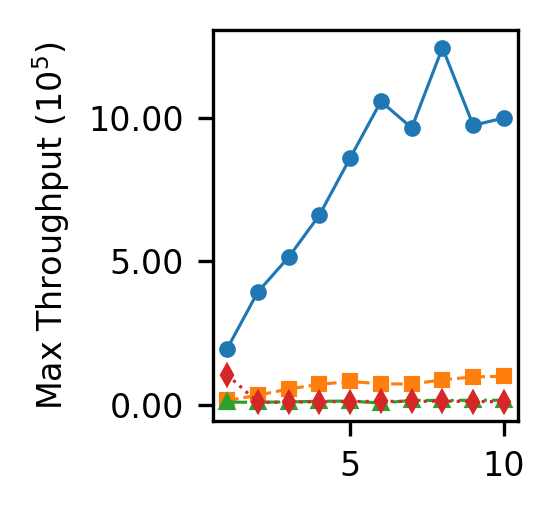}
             \captionsetup{font=scriptsize}
            \caption{Max throughput}
            \label{fig:reddit_hyperlink_throughput_max}
        \end{subfigure}
        \begin{subfigure}{0.24\textwidth}
            \includegraphics[width=\textwidth]{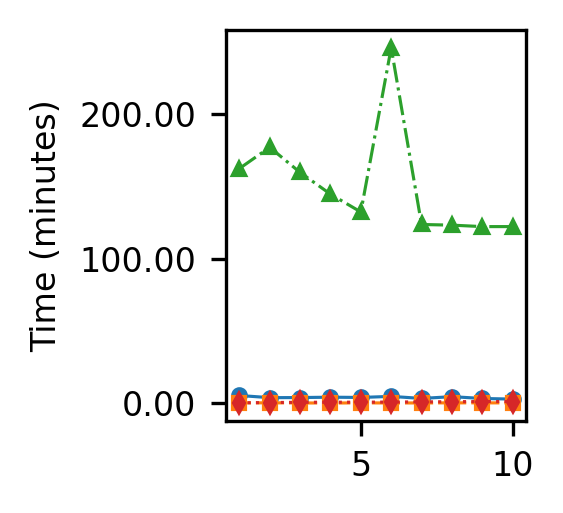}
             \captionsetup{font=scriptsize}
            \caption{Runtime}
            \label{fig:reddit_hyperlink_runtime}
        \end{subfigure}
        \begin{subfigure}{0.24\textwidth}
            \includegraphics[width=\textwidth]{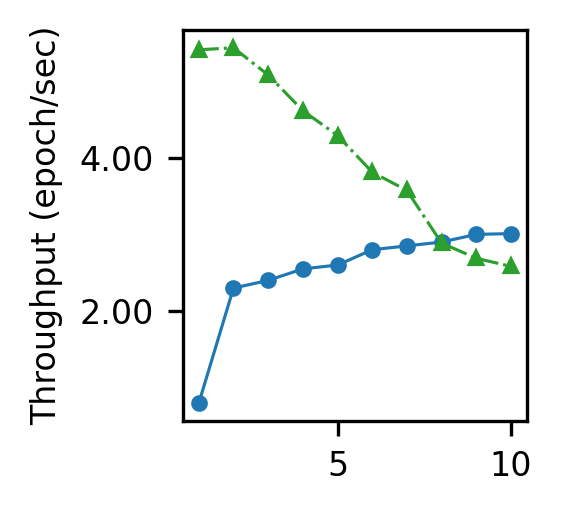}
            \captionsetup{font=scriptsize}
            \caption{Train throughput}
            \label{fig:reddit_hyperlink_training_throughput}
        \end{subfigure}
        \captionsetup{font=small}
        \caption{Comparing inference and training against DGL on \texttt{reddit-hyperlink}}
        
        \label{fig:reddit_hyperlink}
    \end{minipage}
    \hfill
    \begin{minipage}{0.30\textwidth}
        \centering
        \includegraphics[width=0.7\textwidth]{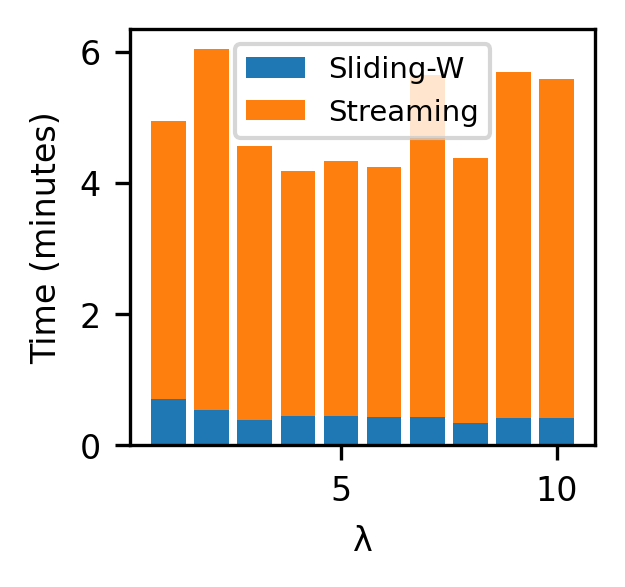}
        \captionsetup{font=small}
        \caption{Explosion factor}
        \label{fig:explosion_factor}
    \end{minipage}

    \Description[Experimental comparison with DGL]{Further experiments indicating performance across DGL, training and explosion factor benefits}
\end{figure*}
\begin{figure*}[ht]
    \centering
    \begin{subfigure}[b]{0.55\textwidth}
        \centering
        \includegraphics[width=0.8\textwidth]{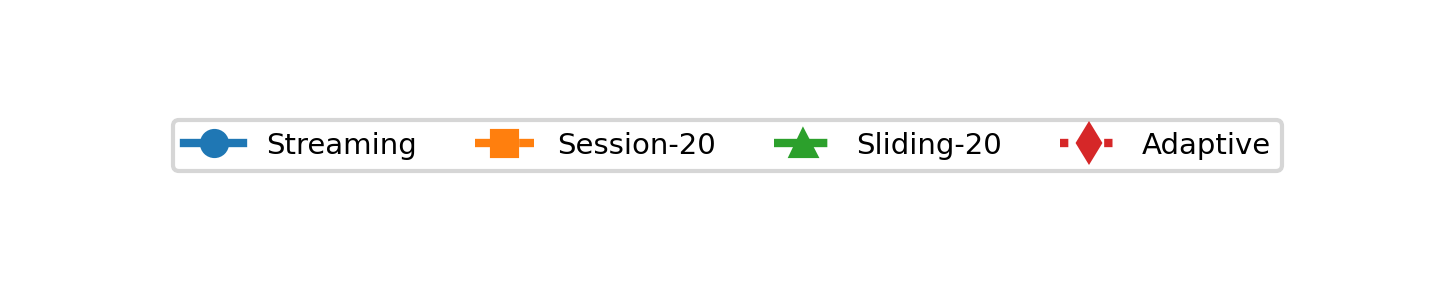}
    \end{subfigure}
    \begin{subfigure}[b]{0.45\textwidth}
        \vspace{-20pt}
        \centering
        \includegraphics[width=\textwidth]{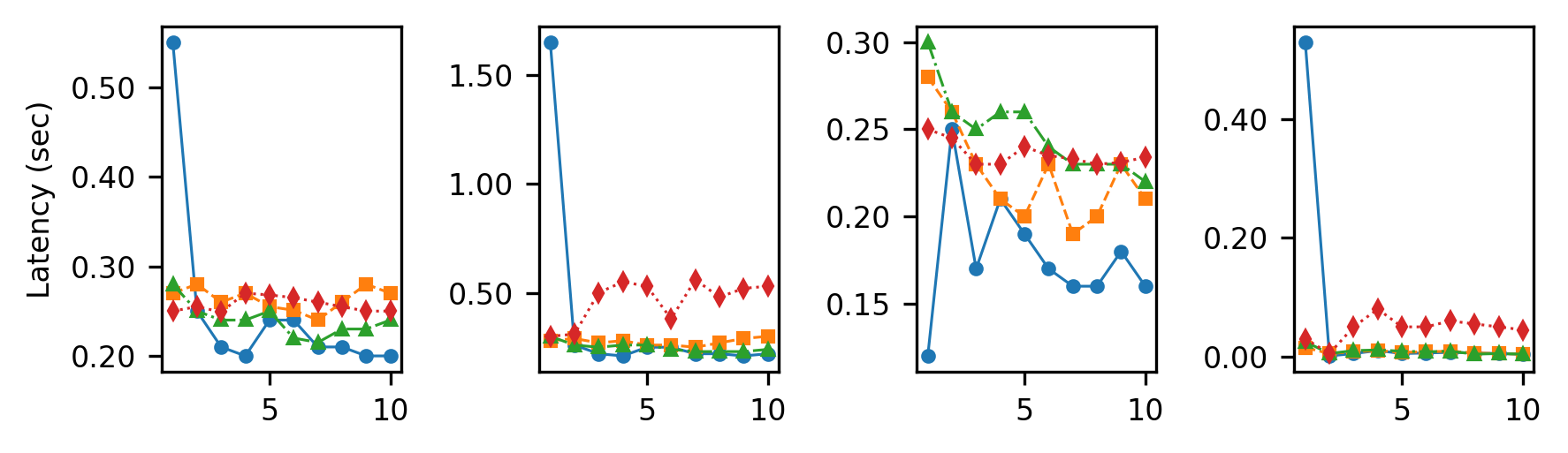}
    \end{subfigure}
    \captionsetup{font=small}
    \caption{Scalability of mean, max, min, and standard deviation of latencies for ingesting 10k edges/sec from \texttt{sx-superuser}} 
    \label{fig:sx-superuser_latency}

    \Description[Latency scalability experiments]{Scalability of mean, max, min, and standard deviation of latencies for ingesting 10k edges/sec from \texttt{sx-superuser}}
\end{figure*}
\noindent\textbf{\textit{Experimental setup and baselines.}}
Experiments are executed on a Slurm cluster with 10 machines, where each machine contains Xeon E5-2660 v3 @ 2.6 GHz (20 cores/40 threads) and 64GB RAM. We use Apache Flink\footnote{\url{https://flink.apache.org}} and Deep Java Library\footnote{\url{https://djl.ai}} with PyTorch\footnote{\url{https://pytorch.org}} as our primary ML framework.

The generated graph representations from the temporal network datasets in the experiments can be further used in a wide range of applications, including fraud detection, social modelling and recommendations by adding an output layer to the final representations. Hence, we focus on generating streaming node representations and build a distributed 2-layer, GraphSAGE model with 64 output dimensions using {\Ours}.
As a baseline, we employ the distributed version of DGL \cite{zheng2020distdgl} with our enhancements to emulate our incremental algorithm described in Section \ref{subsection:gnn_inference}. Since DGL does not support dynamic edge additions, we label each graph edge with a timestamp. For each edge, we simulate topology updates in DGL via a sampling process. That is, our DGL baseline updates the representations for influenced nodes by only sampling from edges prior to its current timestamp. 

We also provide partitioning-based performance evaluation with respect to HDRF, CLDA, METIS and Random vertex-cut partitioners. In HDRF and CLDA we use a balance coefficient of $\theta=2$ and $\epsilon=1$. Furthermore, we have empirically determined the value of $\lambda=3$ as our explosion factor, based on runtime performance.

To evaluate the performance of {\Ours}, we evaluate the following: 
i) Scalability of different inference approaches and partitioners in terms of throughput, running times, network volume and load imbalance; 
ii) Comparison of inference and training with respect to DGL;
iii) Impact of explosion factor on the runtime performance;
iv) Analysing the latency overheads of the windowed inference.
Note that {\Ours} and its streaming incremental aggregators produce the same embeddings as those from a static model executed on the equivalent final graph snapshot, therefore accuracy remains unaffected.

We examine the impact on throughput, total communication volume, latency, load imbalance, and run time when increasing the number of task managers available for allocation. Figure~\ref{fig:sx_superuser} provides introspective analysis of {\Ours}. Whereas, in Figure~\ref{fig:reddit_hyperlink} we compare {\Ours} against DGL using \texttt{reddit-hyperlink}. Figure~\ref{fig:explosion_factor} details the impact of explosion factor on the runtime performance of the system. Lastly, in Figure~\ref{fig:sx-superuser_latency}, we evaluate the latency overheads of windowing on \texttt{sx-superuser}.

We compare 5 types of algorithms - Streaming and Windowed. In \textbf{\Ours\ Streaming} and \textbf{DGL Streaming}, influenced nodes are updated with each incoming edge. 
For {\Ours}, this approach corresponds to Algorithm~\ref{streamingInference} which generates new node embeddings by cascading computation graph updates. In the Windowed case, edges are processed after some delay (20ms except for \texttt{wikikg90Mv2} where delay is set as 10 seconds), that batches these updates. 
We have labeled the windowing algorithms in accordance with the ones in Section~\ref{subsubsection::windowed_forward_pass}. Additionally, to compare against DGL with equal amount of batching, we include \textbf{\Ours\:WCount-2000} and \textbf{DGL WCount-2000}. These, instead process a specific number of edges in a fixed batch size, rather than enforcing a timer-based delay. In our evaluations, we process 2000 edges in a single batch. To ensure consistent throughput as the system scales, we set the window size of each distributed process at ${2000/parallelism}$. 



The reported results are an average of three runs, and the systems were gradually scaled up from 1 to 10 task managers (x-axes were omitted to reduce space). To accurately reflect the performance of the system, we made sure that the entire pipeline was fully busy by congesting it first. 

\noindent\textbf{\textit{Scalability of streaming inference.}}
To measure inference throughput, we calculated the average and maximum rates, across an operator's run time, of producing final layer representations. We observed that {\Ours} scales almost linearly with the number of machines for all algorithms when it comes to throughput performance (Figures~\ref{fig:sx_superuser_throughput_mean},~\ref{fig:reddit_hyperlink_throughput_mean},~\ref{fig:reddit_hyperlink_throughput_max}). However, we also observed that in the case of highly imbalanced graphs, such as \texttt{sx-superuser}, streaming algorithm can suffer at higher levels of parallelism. This is explained by high workload imbalance, which is also reflected in Figure~\ref{fig:sx_superuser_imbalance_factor}. When faced with such circumstances, adopting one of the windowing algorithms has been demonstrated to reduce workload imbalance by nearly $x4$, resulting in enhanced scalability. We calculate imbalance factor by averaging the busy time for each sub-operator and then dividing the maximum by the average. Upon comparing our three windowing algorithms, we find that the adaptive approach yields higher throughput, while the Sliding and Session-based windowing methods are comparable. 

Compared to DGL, {\Ours} demonstrates superior inference task performance, surpassing DGL by a factor of $x76$ on streaming, and by $x15$ on WCount-2000 tasks. Furthermore, we observe that DGL WCount-2000 encounters a significant performance drop when entering distributed mode with two or more machines. These findings are also reflected in the runtime metrics.

\noindent\textbf{\textit{Running time efficiency.}}
When dealing with bounded computations, we use runtime rather than throughput to accurately gauge scalability. To calculate runtime, we utilize the termination detection algorithm explained in Section~\ref{subsection:termination}.

Our study revealed sub-linear scalability in terms of runtime for {\Ours}'s streaming algorithm. Notably, the runtime scalability of windowed approaches are \textbf{super-linear}.  Additionally, for \texttt{sx-superuser} dataset, the streaming algorithm exhibited poor scalability on higher levels of parallelism. This was consistent with the inference throughput, underscoring the advantages of windowing algorithms. Figure~\ref{fig:sx_superuser_runtime} highlights that these algorithms can improve system runtime by a factor of almost 10, while resulting in steeper runtime curves.

Although, DGL WCount-2000 algorithm shows competitive running times in lower parallelisms, for increased parallelisms its runtime consistently worsens. On the other hand, DGL Streaming, which has sub-linear scalability, takes $25x$ of {\Ours} runtime. It is important to highlight the sudden runtime jump of DGL Streaming being run on 6 machines. This was persistent behavior concluded to be due to the partitioning strategy (METIS \cite{metis_partitioning}) for that setting, and was resolved by replacing the partitioner. This further suggests that our asynchronous vertex-cut GNN pipeline provides greater resilience against incorrectly partitioned hub-vertices.

Results for \texttt{stackoverflow} also demonstrate the weak scaling to millions of edges, where the running time is found to decrease following similar trends to that of \texttt{reddit-hyperlink} (58 mins on 5 machines to 20 mins on 10 machines). The figures are omitted due to lack of space.

\noindent\textbf{\textit{Communication volume.}}
Communication operations are incurred due to replication when distributing tasks across a large number of processors. The experiments (Figure~\ref{fig:sx_superuser_comm_volume}) suggest significant reduction in communication volume  (around $15x$) by employing millisecond-scale, windowing algorithms. These measure the volume of iterative communication in the second GNN layer in $GB$. Furthermore, while the streaming communication volume increases sub-linearly with system scaling, the windowing algorithm's communication volume is \textbf{consistent}. This improvement occurs because as the windowing algorithm is able to consume events more rapidly the delay between node updates shortens, hence decreasing the neighborhood explosion. This further supports the significance of our windowing methods for scalable operation. In practice, depending on the network infrastructure, communication volume can be tweaked by using larger windowing intervals.

\balance

\noindent\textbf{\textit{Scalability of training.}}
We conducted a performance comparison of {\Ours} and DGL when training a 2-layer GraphSAGE model for vertex classification (Figure~\ref{fig:reddit_hyperlink_training_throughput}). We did not consider any batching or neighborhood sampling for either system. Despite {\Ours} being designed as a streaming, inference-first system, our results indicate that it performs competitively with DGL. In particular, {\Ours} is more effective at higher scalability, whereas DGL struggles to scale for high-communication workloads.

\noindent\textbf{\textit{Inference latency.}}
Our latency measurements involve determining the time interval between the ingestion of a graph event and the production of the corresponding node representation in the final layer. We included a throttling mechanism to the \texttt{Partitioner} to cap the ingestion rate at $10k$ 
 $edges/sec$. This is necessary to avoid ending in a back-pressured (congested) state. According to our experiments (Figure~\ref{fig:sx-superuser_latency}), when the streaming algorithm can keep pace with the ingestion rate, which it failed to do only for parallelism 1, its latency is the lowest one amongst all algorithms and it shows minimum variance. 

We also observed that adaptive windowing produces a comparable latency to the session one despite having an edge on throughput performance. Overall, {\Ours} achieves sub-second inference latency for at a rate $10k$ $edges/sec$.

\noindent\textbf{\textit{Effect of partitioner.}}
Figure~\ref{fig:sx_superuser} shows HDRF and CLDA surpass Random partitioning across all metrics. Consequently, Random partitioning underperforms, adversely affecting scalability metrics like network volume and runtime. Comparing streaming to METIS static partitioning, static partitioning shows slight superiority. Specifically, METIS significantly cuts network volume in streaming setups. Yet, adopting a windowed approach narrows this performance gap, particularly at higher parallelism levels where network volumes stabilize. This finding suggests that combining HDRF or CLDA with a windowing algorithm can match METIS's performance, a feat not achievable with Random partitioning.

\noindent\textbf{\textit{Effect of explosion factor.}}
Figure~\ref{fig:explosion_factor} demonstrates the beneficial impact of the explosion factor on the \texttt{reddit-hyperlink} dataset. While the theory suggests that the optimal explosion factor should align with the graph's average out-degree, practical outcomes reveal that partitioning schemes, load imbalances, and ingestion order significantly influence performance. For instance, the performance peaks at $\lambda=2$ or $\lambda=7$ can be attributed to load imbalances, as the mapping from logical to physical partitions was not contiguous. Nonetheless, incorporating a windowing approach significantly mitigates these issues, reducing the influence of such factors on overall performance.

\section{Conclusion}

In this work, we introduced {\Ours}, the first distributed, hybrid-parallel system optimized for GNN inference and training in the face of streaming graph updates. Diving into the relatively untouched domain of online query settings for GNNs, {\Ours} excels by incrementally maintaining node embeddings with minimal latency and ensuring fault-tolerant graph data management.
We demonstrated the strong scalability of the system at higher parallelism, with high throughput and low runtime. Furthermore, {\Ours} introduced several algorithmic and systems optimizations, such as improving load balance by intra-layer and inter-layer windowing that reduced runtime by $x10$ and communication by $x7$ in our experiments. Significant improvements over potential alternative designs, including DGL, also highlight the contributions of {\Ours} in handling streaming GNN workloads that require near real-time processing. Lastly, the introduction of a stale-free, synchronous training algorithm by {\Ours} underscores its potential in the machine learning landscape, eliminating the need for separate training environments and addressing bursty resource provisioning challenge. 
\begin{acks}
    This research is supported in part by EPSRC Doctoral Training Partnership
    award (Grant EP/T51794X/1) and Feuer International Scholarship in Artificial Intelligence at University of Warwick.
\end{acks}


\bibliographystyle{ACM-Reference-Format}
\bibliography{sample}

\end{document}